\documentstyle[12pt]{article}
\baselineskip
\advance\textheight by \topskip
\begin{document}
\vspace{-0.2cm}
\begin{flushright}
SU-ITP-95-10\\
CTP \# 2427\\
QMW-PH-95-12\\
hep-th/9505116
\end{flushright}
\vspace{-0.5cm}

\begin{center}
{\Large\bf
Fermion Zero Modes and Black-Hole Hypermultiplet\\
\vskip 0.5truecm
with Rigid Supersymmetry
}
\vskip 0.8 cm
{
{\Large R}{OGER} {\Large B}{ROOKS}\footnote{On leave
from the Physics Department, Massachusetts Institute of Technology,
Cambridge, MA 02139, U.S.A.},\ \ \  \
{\Large R}{ENATA} {\Large K}{ALLOSH}
}

{\it{Physics Department,
Stanford University, Stanford, California 94305, U.S.A.}}

{and}

{
{\Large T}{OM\'{A}S} {\Large O}{RT\'{I}N}\footnote{Address after
October $1^{st}$ $1995$: Theory Division, CERN, CH--1211, Gen\`{e}ve
23, Switzerland}
}

{\it{Physics Department,
Queen Mary and Westfield College}}\\
{\it{Mile End Road, London E1 4NS, U.K.}}
\end{center}

\begin{abstract}

The gravitini zero modes riding on top of the extreme
Reissner--Nordstr\"om black-hole solution of $N=2$ supergravity are
shown to be normalizable.  The gravitini and dilatini zero modes of
axion-dilaton extreme black-hole solutions of $N=4$ supergravity are
also given and found to have finite norms.  These norms are duality
invariant. The finiteness and positivity of the norms in both cases

are found to be correlated with the Witten--Israel--Nester
construction;   however, we have replaced the  Witten condition  by the
pure-spin-${3\over 2}$ constraint on the gravitini.
We compare our calculation of the norms with the calculations which
provide the moduli space metric for extreme black holes.

The action of the $N=2$ hypermultiplet with an off-shell central
charge
describes the solitons of $N=2$ supergravity.  This action, in the
Majumdar--Papapetrou multi-black-hole background, is shown to be $N=2$
rigidly supersymmetric.

\end{abstract}

\newpage
%%%%%%%%%%%%%%%%%%%%%%%%%%%%%%%%%%%%%%%%%%%%%%%%%%%%%%%%%%%%%%%%%%%%%%

%

\section{Introduction}
%{{\hskip 10pt\vskip -15pt}}

Classical solutions of supergravity theories with unbroken
supersymmetries have attracted much attention in recent years.  This
is
due to the many interesting properties that they usually share: they
minimize the energy for given values of the charges (they saturate
supersymmetric Bogomol'nyi bounds \cite{Bog}) and so they are stable
and
can be considered solitons \cite{Raj}, alternative vacuums or ground
states of the theory (depending on the number of unbroken
supersymmetries).  In general, multi-center solutions describing an
arbitrary number of these solitons in equilibrium exist.  In some
cases
they also enjoy non-renormalization theorems \cite{Kal2}.

The supersymmetric solutions usually considered have vanishing
fermionic
fields.  It is, however, easy to obtain new solutions with
non-vanishing fermion fields starting with the purely bosonic ones
and
performing supersymmetry transformations.  If the supersymmetry
parameters used vanish asymptotically ({\it i.e.}~the supersymmetry
transformations are {\it trivial}), the new solutions obtained in
this
way are gauge-equivalent to the original ones.  If the supersymmetry
parameters converge asymptotically to global supersymmetry parameters
then the new solutions obtained are no longer gauge-equivalent to the
original ones (it is not possible to go back to them by using
asymptotically vanishing supersymmetry parameters).  In this way, if
non-trivial supersymmetry parameters with the right regularity
properties exist, one can generate a whole supermultiplet of
solutions
\cite{Gibb}.  This program has been successfully carried out for the
Majumdar--Papapetrou solutions of $N=2$ supergravity in a series of
papers by Aichelburg, {\it et al.}~\cite{Aich}.

{}From the point of view of the representation theory of the
supersymmetry
algebra \cite{FerSavZum}, the supermultiplets of solutions generated
in
this way are {\it shortened} supermultiplets whose dimension is
smaller
than the dimension of the original supersymmetry multiplet.
This is so because, by definition, there are some non-trivial
supersymmetry transformations (those generated by the Killing
spinors)
that leave the original solution invariant.  Only the non-trivial
supersymmetry parameters corresponding to the broken supersymmetries
(that we will call ``anti-Killing spinors") generate new non-gauge
equivalent solutions.

Having found the supermultiplet structure of these supersymmetric
solitons, it is natural to look for a theory describing its dynamics.
This would be the supersymmetric quantum field theory of the solitons
of
the original theory.  There are well-known examples of quantum field
theories which describe the quantum relativistic dynamics of the
solitons of another theory.  The most famous example of this {\it
duality} is the relation between the Thirring model and the
sine-Gordon
model.  The former has as elementary excitations the solitons of the
second one \cite{Col}.  It was argued by Montonen and Olive
\cite{MonOli} that there should exist quantum field theories of the
magnetic
monopoles of known gauge theories.  An example along the lines of
this
conjecture was pointed out by Osborn in Ref.~\cite{Osb}.  He showed
that
the
spectrum of solitons of $d=4,\, N=4$ super-Yang--Mills, which have
two
unbroken supersymmetries, is the same as the spectrum of elementary
excitations of the original theory and suggested that the theory
could
be
self-dual.

Although the Montonen--Olive conjecture does not directly apply to
the
kind of solitons that we will be concerned with here (extreme black
holes), it is possible that a quantum field theory describing the
dynamics of these objects exists.  The supermultiplet structure
determines, to a large extent, the form of the theory.  In
particular,
since black holes would correspond to massive states, massive
representations of the supersymmetry algebra are needed.  The
supergravity multiplets one starts with are massless, and, therefore,
they cannot describe the supermultiplets of black-hole solutions.

Another property of the purely bosonic supersymmetric solutions is
that
they admit fermion zero modes, {\it i.e.}~classical solutions of the
fermion (Dirac or Rarita--Schwinger) equations of motion in that
background.  Furthermore, it is straightforward to generate these
fermion zero modes: an infinitesimal supersymmetry transformation
generated by anti-Killing spinors generates non-trivial fermion
fields
which solve the equations of motion and do not change the bosonic
background.
The fermion zero modes are thus the lowest order contribution to
the fermion fields in each solution in the
supermultiplet.

So far, the normalizability of these fermion zero modes (which is an
important issue if the supermultiplet of solutions is going to be
interpreted as a supermultiplet of states) had always been implicitly
assumed.  It has recently been shown in Ref.~\cite{BecStr} that this
is
not always so.  This raises the question as to whether in the
previously
known cases the norms of the fermion zero modes (which never were
explicitly
calculated) are finite or not, and why.

In this paper we will demonstrate the existence of normalizable
fermion
zero modes in the extreme Reissner--Nordstr\"om (ERN) black-hole
background and in the extreme dilaton black-hole (EDBH) background.
We
will show that the normalizability of these zero modes can be
understood
in terms of the saturation of the Bogomol'nyi positivity bounds for
the
ADM mass by using the four--dimensional $N=2$ and $N=4$ versions of
the
Witten--Israel--Nester (WIN) \cite{IsNes} construction presented in
Refs.~\cite{GibbHull} and \cite{HarvLiu}, respectively.

In addition, we will investigate the action of the massive $N=2$
hypermultiplet whose supermultiplet structure corresponds to that of
the
ERN multiplet and it is therefore {\it the} candidate to describe the
dynamics of the ERN black holes.  We will see that the existence of
fermionic zero modes in the ERN black-hole background leads to the
existence of rigidly (as opposed to globally) supersymmetric
theories.
In particular, we will see that the $N=2$ hypermultiplet may be
placed
on an ERN background.  We expect that the techniques used here may be
applicable to other curved geometries.

%%%%%%%%%%%%%%%%%%%%%%%%%%%%%%%%%%%%%%%%%%%%%%%%%%%%%%%%%%%%%%%%%%%%%%

%

\section{Normalizability of the ERN zero modes}
%{{\hskip 10pt\vskip -15pt}}

To begin, we recall some well-known facts.  We start by describing
the
Majumdar--Papapetrou (MP) solutions of the Einstein--Maxwell theory
\cite{M-P} for zero magnetic and positive electric
charges\footnote{Our
metric's signature is $(+---)$, the gamma matrices are those of
Ref.~\cite{AichUrb} and in particular satisfy
$\{\gamma^{\mu},\gamma^{\nu}\}= +2g^{\mu\nu}$, where $\gamma_{0}$ is
Hermitean, the $\gamma_{i}$'s are anti-Hermitean and $\gamma_{5} =
i\gamma_{0}\gamma_{1}\gamma_{2}\gamma_{3}$.  Greek indices are
curved,
the first Latin alphabet letters $a,b,c,\dots$ are flat indices and
the
indices $i,j,k,\ldots$ run from $1$ to $3$.  Underlined indices
($\underline{0}$,$\underline{i}$, etc.) are always curved.
$\epsilon^{\underline{0}\underline{1}\underline{2}\underline{3}}
=+(-g)^{-\frac{1}{2}}$.},

\begin{equation}
ds^2~=~ V^{-2} dt^2 ~-~ V^2 d{\vec x}^2\ \ ,\qquad
A~=~ -{1\over{\kappa}} V^{-1}~dt\ \ ,
\end{equation}

\noindent where $\kappa^{2}=4\pi G$ and the function $V$ does not
depend on time and satisfies

\begin{equation}
\partial_{\underline{i}}\partial_{\underline{i}}V=0\, .
\label{Laplace}
\end{equation}

\noindent The requirements of asymptotic flatness and regularity
determine it to be of the form\footnote{Dropping the condition of
asymptotic flatness other solutions are possible.  Remarkably enough,
if
one deletes the $1$ in Eq.~(\ref{V}) we get, for a single
$\vec{x}_{s}$, Robertson--Bertotti's solution.}

\begin{equation}
V(\vec{x})\ =\ 1\ +\ \sum_{s} \frac{G M_s}{|\vec x\ -\ \vec x_s|}\, ,
\label{V}
\end{equation}

\noindent where the horizon of the $s$th black hole is located at
$\vec{x}=\vec{x}_{s}$ and its electric charge is
$Q_{s}=G\kappa^{-1}M_{s}$.  The parameter $M_{s}$ is usually
interpreted
as the mass of the $s$th black hole.  However, in this space-time,
there
is no way to calculate the mass of each individual black hole since
there is only one asymptotically flat region, common for all black
holes.  Thus, strictly speaking, one can only say that the ADM mass
of
this space-time is $M_{ADM}=\sum_{s}M_{s}$.  Nevertheless, taking
into
account that this background describes charged black holes in static
equilibrium which share many of the characteristics of ERN black
holes,
it is natural to identify $M_{s}$ with the mass of the $s$th black
hole.

Since the multiplet of $d=4, N=2$ supergravity \cite{FerNieu,SFZ} is
$(e_\mu{}^a, A_\mu, \psi_\mu)$, where $A_\mu$ is a $U(1)$ gauge field
and $\psi_\mu$ is a complex spin--$\frac{3}{2}$ field ({\it i.e.}~it
is
a Dirac spinor for each value of the vector index $\mu$), it is clear
that the Einstein--Maxwell theory can be embedded in it, and the MP
solutions can be considered solutions of $d=4,\, N=2$ supergravity
with
the only fermion field of the theory vanishing: $\psi_{\mu}=0$.

A remarkable feature of this background is that it admits $N=2$
supergravity Killing spinors \cite{GibbHull,Tod}, {\it i.e.}~a
solution
of the equation

\begin{equation}
\hat\nabla_\mu
\epsilon~=~0\ \ .
\end{equation}

\noindent Here $\hat\nabla_\mu$ is the $N=2$ supercovariant
derivative
given in terms of the gravitational covariant derivative,
$\nabla_\mu$ by
\begin{equation}
\hat\nabla_\mu ~\equiv~ \nabla_\mu -\frac{1}{4} \kappa\ {{\slash
\hskip -8pt}{} F}\gamma_\mu\ \ ,
\end{equation}
where ${{\slash \hskip -8pt}{}
F}=\gamma^{\mu\nu} F_{\mu\nu}$ and $F_{\mu\nu}$ is the
field--strength
of the gauge field $A_\mu$ and $\epsilon$ is a Dirac spinor.  That
solution
is given by

\begin{equation}
\epsilon_{(k)} ~=~ V^{-1/2}{\cal C}_{(k)}\ \ ,
\end{equation}

\noindent where ${\cal C}_{(k)}$ is a constant spinor satisfying the
condition

\begin{equation}
\gamma_0\ {\cal C}_{(k)} = +\ {\cal C}_{(k)}\, ,
\label{property}
\end{equation}

\noindent and is given in terms of a complex two-component spinor,
$c$,
by ${\cal C}_{(k)}{}^t = (c^{t},c^{t})$.  This means that the
background
given above has one unbroken supersymmetry in $N=2$ supergravity.

If we perform an infinitesimal supersymmetry transformation with a
supersymmetry parameter $\epsilon$ that does not vanish
asymptotically
and that it is not a Killing spinor either (we will call it an
``anti-Killing" spinor and denote it by $\epsilon^{(\bar k)}$), the
bosonic fields (the metric and vector field) will remain invariant
but a
non-trivial fermionic field that solves the gravitini field equations
in
this background will be generated.  This fermionic zero-mode is
therefore given by \cite{Aich}

\begin{equation}
\psi_\mu~=~ {1\over{\kappa}} \hat\nabla_\mu\epsilon^{(\bar k)}\ \ ,
\label{GRAV0M}
\end{equation}

\noindent where $\epsilon^{(\bar k)}$ is the anti-Killing spinor.
By construction it satisfies the covariant Dirac equation

\begin{equation}
\gamma^\mu\hat\nabla_\mu \epsilon^{(\bar k)}~=~
\gamma^\mu\nabla_\mu \epsilon^{(\bar k)}~=~0\ \ ,\label{Dirac}
\end{equation}

\noindent which implies that the gravitini constructed in this way is
always in the gauge in which

\begin{equation}
\gamma^{\mu}\psi_{\mu} ~=~ 0\, .
\end{equation}

In our case the anti-Killing spinor can be chosen to be, in terms of
the Killing spinor,

\begin{equation}
\epsilon^{(\bar k)} = -i\gamma_5 \epsilon_{(k)}\, ,
\label{anti-Killing}
\end{equation}

\noindent so the explicit expression for the gravitino zero-mode is

\begin{eqnarray}
\label{n=2grav}
\psi &~=~& {1\over{\kappa}} V^{-7/2} \partial_{\underline{i}} V
\gamma_i
{\cal C}^{(\bar k)}~dt~+~
{1\over{\kappa}} V^{-3/2} \partial_{\underline{j}} V \gamma_j
\gamma_i
{\cal C}^{(\bar k)}~dx^{\underline{i}}\ \ ,\nonumber \\ \relax
& & \nonumber \\
{\cal C}^{(\bar k)}& ~=~ & {c\choose{-c}}\ \ .
\end{eqnarray}
Observe that the property  of the Killing spinor given in
Eq.~(\ref{property})
implies that

\begin{equation}
\gamma_0\ {\cal C}^{(\bar k)} = -\ {\cal C}^{(\bar k)}\, .
\label{antiproperty}
\end{equation}

If we performed a finite supersymmetry transformation generated by
the
anti-Killing spinor $\epsilon^{(\bar k)}$, Eq.~(\ref{n=2grav}) would
be
the lowest order, in $\epsilon^{(\bar k)}$, contribution to the
gravitino.  The normalizability of the zero modes is then related
to
the question of whether a (normalizable) supermultiplet of solutions
can
be built starting from the MP solutions.

The norm of the gravitino is, by definition

\begin{equation}
\|\psi\|^{2}~\equiv~\int_\Sigma d^3x \sqrt{-g_{(3)}}~
{\psi}_\mu{}^\dagger\psi_\nu g^{\mu\nu}\ \ ,
\label{NORM2a}
\end{equation}

\noindent where $\Sigma$ is a space-like hypersurface and $g_{(3)}$
is
the determinant of the induced metric on it.  In our case, $\Sigma$
will
be any constant-time hypersurface and $g_{(3)}={\rm det}
g_{\underline{i}\underline{k}}$.  The norm  of the zero-mode
of
Eq.~(\ref{n=2grav}) is

\begin{equation}
\|\psi\|^{2} {}~=~
{1\over{2\pi G}}{\bar{\cal C}}^{(\bar k)}{\cal C}^{(\bar k)}
\int_{D} d^3x~ V^{-2}(\partial_{\underline{i}} V)^2\ \ .
\end{equation}

\noindent where the integration domain $D$ is a subset of the
three-dimensional ${{\hbox{I}\hskip -3.3pt\hbox{R}}\hskip -2pt}^{3}$
to be
determined later.  Let us now
specialize to the single ERN black-hole background.  In this case $D$
is
${\hbox{I}\hskip -3.3pt\hbox{R}\hskip -2pt}^{3}$ with the origin
removed
(${\hbox{I}\hskip -3.3pt\hbox{R}\hskip -2pt}^3{-\{\vec{0}\}}$) and we
find by a
direct calculation of the volume integral that\footnote{As a matter
of
fact, we are integrating over a constant time--slice of the ERN
geometry.  It is well known that these hypersurfaces become
bottomless
tubes when one approaches the horizon, which is at infinite proper
distance over this hypersurface and therefore is not even included in
it.}

\begin{equation}
\|\psi\|^2~=~ 2M
{\bar{\cal C}}^{(\bar k)}{\cal C}^{(\bar k)}~=~ 4M \|c\|^2\ \ ,
\label{NORM2b}
\end{equation}

\noindent where $\|c\|^2$ is the norm of the complex, two-component
constant spinor.  Thus the spin--$\frac{3}{2}$ zero-mode in the
ERN black hole background
is
normalizable.

In order to extend this result to the multi-black-hole case, we first
observe that the integrand of the norm is well behaved everywhere
(including the origin of ${\hbox{I}\hskip -3pt\hbox{R}\hskip
-2pt}^{3}$, which
corresponds to the horizon):

\begin{equation}
\int_{D} d^3x~ V^{-2}(\partial_{\underline{i}} V)^2
{}~=~
4\pi G^2 M^2 \int_0^\infty dr  {1\over{(r+GM)^2}} \ \ ,
\end{equation}

\noindent so we do not expect singular contributions to the integral.
In fact, we could have used Gauss' theorem to evaluate the norm as a
surface integral at infinity in ${\hbox{I}\hskip -3pt\hbox{R}\hskip
-2pt}^{3}$.
To do this, we first
rewrite
the integrand

\begin{equation}
V^{-2}(\partial_{\underline{i}} V)^2 = -\partial_{\underline{i}}
(V^{-1}\partial_{\underline{i}}V) + V^{-1}
\partial_{\underline{i}} \partial_{\underline{i}}V\, .
\end{equation}

\noindent The second term does not contribute to the integral because
it
vanishes everywhere outside the horizon
(Eqs.~(\ref{Laplace},\ref{V}))
and on the horizon it appears multiplied by a factor which vanishes
there. We get

\begin{equation}
\int_{D}d^3x~ V^{-2}(\partial_{\underline{i}} V)^2 ~=~ -
\int_{D} d^3x~
\partial_{\underline{i}}(V^{-1}\partial_{\underline{i}} V) ~=~ -
\int_{\partial D}
dS^{\underline{i}}\ V^{-1}\partial_{\underline{i}} V\, ,
\end{equation}

\noindent where we have applied Gauss' theorem and, in the last
expression the index $\underline{i}$ is a covariant index in $\Sigma$
so
we can perform the surface integral, in particular, in spherical
coordinates.
%Since $D={\hbox{I}\hskip -3.3pt\hbox{R}}^{3}_{-\{0\}}$,
Accordingly, the boundary
consists of two disconnected pieces, one at infinity
($S_{\infty}^{2}$)
and the origin (the horizon).  The surface integral at the horizon
vanishes because the integrand vanishes there (no
singularities there) and we have

\begin{equation}
\int_{D}d^3x~ V^{-2}(\partial_{\underline{i}} V)^2 ~=~ -
\int_{S_{\infty}^{2}}
dS^{\underline{i}}\ V^{-1}\partial_{\underline{i}} V\, .
\end{equation}

\noindent (Alternatively, knowing that there are no singular
contribution to the integral there, we could have included the origin
in
$D$ and $\partial D=\partial {\hbox{I}\hskip -3.3pt\hbox{R}\hskip
-2pt}^{3}
=S_{\infty}^{2}$, getting the
same
final result.)  The resulting surface integral is easy to calculate
and
gives the expected result.

Now it is clear that exactly the same arguments go through in the
multi-black-hole case ($D={\hbox{I}\hskip -3.2pt\hbox{R}\hskip
-2pt}^{3}{-\{\vec{x}_{a}\}}$), where the
volume
integral becomes too complicated, and we obtain a surface integral
which
is asymptotically identical to the one in the single-black-hole case.
To summarize, for any number of black holes, the norm of the
gravitini
zero modes is given by

\begin{equation}
\|\psi\|^{2} {}~=~ -
{1\over{2\pi G}}{\bar{\cal C}}^{(\bar k)}{\cal C}^{(\bar k)}
\int_{S_{\infty}^{2}}
dS^{\underline{i}} \, V^{-1}\partial_{\underline{i}} V {}~=~
4M_{ADM}\| c \|^{2}\, ,
\end{equation}

\noindent where $M_{ADM}=\sum_{s}M_{s}$ is the total ADM mass.

We would like to understand what is the underlying reason for the
normalizability of the gravitini zero modes, since it seems to be
just
``pure luck'' that they are normalizable in some cases and
not normalizable in others \cite{BecStr}.  Before proceeding we note
that the fact that the norm can rewritten as a surface integral over
$S^2_\infty$ is suggestive of an ADM construction.

The first crucial observation is that the norm of the gravitino
zero-mode,
as defined in Eq.~(\ref{NORM2a}) is the starting point in the
$N=2$ generalization \cite{GibbHull} of the WIN construction
\cite{IsNes}.  Indeed, since the gravitino is obtained by a
supersymmetry
transformation as in Eq.~(\ref{GRAV0M}) with a spinor $\epsilon$,
we can proceed as follows.  First we define

\begin{equation}
I~\equiv~\frac{1}{\kappa^{2}}
\int_\Sigma d\Sigma_\mu\
\overline{\hat{\nabla}_{\nu}\epsilon}\ \gamma^{\mu\nu\rho}
\hat{\nabla}_{\rho}\epsilon\ \ .
\end{equation}
Then using Eq.~(\ref{GRAV0M}) we see that for our gravitini,
\begin{equation}
I~=~\int_\Sigma d\Sigma_\mu\ {\bar\psi_\nu}\gamma^{\mu\nu\rho}
\psi_\rho\ \ .
\label{first}
\end{equation}

%\noindent and, making the same choice of hypersurface $\Sigma$
\noindent Next, it follows that

\begin{eqnarray}
\int_\Sigma d\Sigma_\mu\ {\bar\psi_\nu}\gamma^{\mu\nu\rho}\psi_\rho
&~=~& \int_\Sigma d^3x \sqrt{-g_{(3)}}\
{\psi_j}^\dagger\gamma^0\gamma^{0jk}\psi_k\ \ ,
\nonumber \\
\relax
& & \nonumber \\
&~=~&  \int_\Sigma d^3x \sqrt{-g_{(3)}}\ \psi_\mu^\dagger \psi_\nu
g^{\mu\nu}~=~ \|\psi \|^{2}\ \ .
\end{eqnarray}

\noindent It is noteworthy that in deriving this relation for the gravitino
norm, we did  not
use the
Witten condition, $\gamma^{\underline i}{\hat \nabla}_{\underline i}\epsilon=
0$, which is standard in
the WIN construction when deriving the Bogomol'nyi bound. Instead, we have
imposed the condition that the gravitino is a pure
spin-${3\over 2}$ field: namely,
\begin{eqnarray}
\gamma^\mu\psi_\mu &~=~& 0\qquad\Longrightarrow\qquad
\gamma^\mu{\hat\nabla}_\mu\epsilon~=~0\ \ ,
\end{eqnarray}
Observe that for the MP configuration these equations  imply
\begin{equation}
\gamma^{\underline i}\psi_{\underline i}~\neq~0\qquad\Rightarrow\qquad
\gamma^{\underline i}{\hat \nabla}_{\underline i}\epsilon~\neq~0\ \ .
\end{equation}

Continuing with our analysis, we see that
the integral $I$ in Eq.~(\ref{first}), for any
spinor $\epsilon$ that asymptotically approaches the constant spinor
${\cal C}$ ($\epsilon={\cal C}+ O(1/r)$) is equal to \cite{GibbHull}

\begin{equation}
I~=~\int_{\Sigma}d\Sigma_{\alpha}\
\overline{\cal C}\left[T^{(matter)\,\alpha}{}_{\beta}\gamma^{\beta}
+\kappa G^{-1}\left(J^{\alpha} +i\gamma_{5}\tilde{J}^{\alpha} \right)
\right]{\cal C}
-\overline{\cal C}\left[
-P_{\lambda}\gamma^{\lambda}+\kappa G^{-1}(Q+i\gamma_{5}P)\right]
{\cal
C}\, .
\label{lhs}
\end{equation}

Let us now specialize these expressions to the case at hand,
$\epsilon=\epsilon^{(\bar k)}$, zero magnetic charge, etc.  To begin
with,
the first term in Eq.~(\ref{lhs}) is identically zero, since all
the sources vanish outside the horizon.  Secondly, one can check that
the only non-vanishing component of the Lorentz vector
$\overline{\cal
C}^{\bar k}\gamma^{a}{\cal C}^{\bar k}$, has $a=0$.  Finally, given
the
property (\ref{antiproperty}) and upon using the relation
$Q=G\kappa^{-1}\sum_{s}M_{s}$
between the total charge  and the ADM
mass
$M_{ADM}$ of the MP solutions, Eq.~(\ref{lhs}) yields :

\begin{equation}
I~=~\left(P^{0}+\kappa G^{-1}Q\right)
({\cal C}^{\bar k})^{\dagger} {\cal C}^{\bar k}
=2M_{ADM}({\cal C}^{\bar k})^{\dagger} {\cal C}^{\bar k}\, .
\end{equation}

\noindent Hence we see that the results of Ref.~\cite{GibbHull} lead
us
directly to the norm of the gravitino.  Our explicit evaluation,
c.f.~Eqns.~(\ref{NORM2a}-\ref{NORM2b}), of the norm is to be viewed
as a
verification of this result.

Now we would like to extend our results to magnetically charged black
holes and dyons which can be obtained by electric-magnetic duality
rotations of the electromagnetic field strength $F^{\mu\nu}$ in the
Einstein--Maxwell theory.  The generalization of this symmetry of the
equations of motion of the Einstein--Maxwell theory to $N=2$
supergravity is known from the early days of the theory \cite{SFZ}
and
is the so-called ``chiral-dual'' symmetry.  The finite chiral-dual
transformations of the gravitino and the supercovariant
electromagnetic
tensor $\hat{F}^{\mu\nu}$  are

\begin{equation}
\psi_{\mu}^{\prime} = e^{\frac{i}{2}\theta \gamma^{5}} \psi_{\mu}\, ,
\hspace{1cm}
\hat{F}_{\mu\nu}^{\pm\prime} = e^{\pm i\theta}
\hat{F}_{\mu\nu}^{\pm}\ \
,
\end{equation}

\noindent where $\hat{F}_{\mu\nu}^{\pm}$ is the (anti-) self-dual
part
of $\hat{F}$:

\begin{equation}
\hat{F}^{\pm}_{\mu\nu}=
{\textstyle\frac{1}{2}}\left(\delta^{\rho\sigma}_{\mu\nu}
\pm {\textstyle\frac{i}{2}}\epsilon_{\mu\nu}{}^{\rho\sigma}\right)
\hat{F}_{\rho\sigma}\, .
\end{equation}

We have calculated the gravitino norm for the electrically charged
black
hole.  Instead of performing the new calculations for the magnetic
one
or for the electromagnetic one, we have only to use the symmetry of
the
norm
(\ref{NORM2a}) under chiral-dual rotation of the vector field and
gravitino:

\begin{equation}
\left( \psi_{\mu}{}^{\dagger}\psi_{\nu} \right)^{\prime} =
\psi_{\mu}{}^{\dagger}\psi_{\nu}\, .
\end{equation}

Alternatively one could consider the transformation rule of $N=2$
Killing spinors under duality:

\begin{equation}
\epsilon_{(k)}^{\prime} = e^{\frac{i}{2}\theta \gamma^{5}}
\epsilon_{(k)}\, ,
\end{equation}

\noindent which implies that our anti-Killing spinors
(\ref{anti-Killing}) and gravitino zero modes transform in the same
way.  Once again, the
norm of the gravitino is duality invariant.

%%%%%%%%%%%%%%%%%%%%%%%%%%%%%%%%%%%%%%%%%%%%%%%%%%%%%%%%%%%%%%%%%%%%%%

\section{Normalizability of the dilaton black-hole zero modes}
%{{\hskip 10pt\vskip -15pt}}

It is interesting to see whether the same happens in other cases. The
simplest extension is  the purely electric, extreme dilaton black
holes (EDBH) \cite{GMGHS} which have two unbroken supersymmetries
when
embedded in $d=4,\ N=4$ supergravity \cite{5auth}.  The fields of the
electric EDBH are
\begin{eqnarray}
ds^2 & = &  V^{-1} dt^2 ~-~ V d{\vec x}^2\, ,
\nonumber \\
& & \nonumber \\
A & = & -\frac{e^{+\kappa \phi_{0}}}{\kappa\sqrt{2}} V^{-1}~dt\, ,
\nonumber \\
& & \nonumber \\
e^{-2\kappa \phi} & = & e^{-2\kappa \phi_{0}} V\, ,
\end{eqnarray}

\noindent where $V$ is given\footnote{As different from the ERN case,
dropping the $1$ in the expression for $V$ does not give another
solution.} by an expression similar to Eq.~(\ref{V}):

\begin{equation}
V(\vec{x})\ =\ 1\ +\ \sum_{s}\frac{2G M_s}{|\vec x\ -\ \vec x_s|}\, .
\label{V2}
\end{equation}

\noindent The {\it mass} of the $s$th EDBH is $M_{s}$, its
electric
charge is $Q_{s}=\sqrt{2}e^{+\kappa \phi_{0}}G \kappa ^{-1}M_{s}$ and
its dilaton charge is $\Sigma_{s}=-\frac{e^{-2\kappa \phi_{0}}
Q_{s}^{2}
}{2G\kappa ^{-1}M_{s}} =-G\kappa ^{-1}M_{s}$ and the $d=4,\ N=4$
supergravity Bogomol'nyi bound is saturated for each black hole:

\begin{equation}
M_{s}^{2} + \kappa ^{2}G^{-2}\left(\Sigma_{s}^{2}-e^{-2\kappa
\phi_{0}}Q_{s}^{2}
\right)=0\, .
\end{equation}

We are interested in only establishing the finiteness of the norm of
the
gravitini and dilatini.  The values of these norms will depend on the
coefficients these fields appear with in the $N=4$ supergravity
action.
Since these numbers will be convention--dependent, we will simply
write
our expressions in terms of the norms of the constant spinors these
zero modes are given in terms of.  In particular, the gravitini and
dilatini zero modes are:

\begin{eqnarray}
\psi_{I} & = &  {1\over \kappa } [\sigma_{0i}
V^{-9/4}\partial_{\underline{i}} V dt
{}~-~ 2 V^{-5/4}(\partial_{\underline{i}} V
-2\sigma_{ij}\partial_{\underline{j}} V)]
{\cal C}_{ I}^{(\bar k)} dx^i \ \ ,
\nonumber \\
& & \nonumber \\
\lambda_{I} & = & \frac{1}{\kappa } V^{-\frac{7}{4}} \
\gamma_{i} \partial_{\underline{i}} V {\cal C}_{I}^{(\bar k)}\, .
\end{eqnarray}

\noindent The norms of these fields are then found to be given by a
hypersurface integral times the norms, $\|{\cal C}_I^{(\bar k)}\|^2$,
of
the constant, Majorana spinors, ${\cal C}_I^{(\bar k)}$.  In terms of
$V$, this hypersurface integral is the same as that which appeared in
the
$N=2$ case.  As we saw above, functionally, the $V$'s differ only in
that the mass, $M$, in the $N=2$ case is replaced by $2M$ for $N=4$.
Hence the calculation of the $N=4$ norms follows from the $N=2$ case.
We then find

\begin{equation}
\|\psi_{I}\|^{2} {}~=~  M\| {\cal C}_I^{(\bar k)}\|^2 \ \ ,
\end{equation}

\noindent and

\begin{equation}
\|\lambda_{I}\|^{2} {}~=~  M\| {\cal C}_I^{(\bar k)}\|^2\ \ .
\end{equation}

\noindent Thus both the gravitini and dilatini zero modes are
normalizable.

That these fields have finite norm also follows from the Nester
theorem
for $N=4$ supergravity \cite{HarvLiu}.  However, explicit values for
the
individual norms of the gravitini and dilatini cannot be obtained
from
that construction.  As before, we have not used the Witten condition.

We can use $S$-duality for the evaluation of the norm of the
gravitino
and
dilatino, for axion-dilaton black holes \cite{ADBH} with axion and
dilaton fields $a(x)$ and $e^{-2\phi (x) }$, which can be generated
out
of the purely electric ones that we have considered above.
According to the discussion of the $N=2$ case, it is enough to know
how
the supersymmetry transformation rules of the fermions behave under
$SL(2,{\hbox{I}\hskip -3.5pt\hbox{R}})$
transformations\footnote{Quantum
mechanical effects
break this symmetry group to $SL(2,Z)$.}
\cite{TO}

\begin{equation}
(\delta_{\epsilon} \psi_{I\mu})^{\prime}
=e^{\frac{i}{2}\gamma^{5}Arg(S)} \delta_{\epsilon} \psi_{I\mu} \ ,
\qquad
(\delta_{\epsilon} \lambda)_{I}^{\prime}
=e^{\frac{3i}{2}\gamma^{5}Arg(S)} \delta_{\epsilon}\lambda_{I} \ .
\end{equation}

\noindent where $S\equiv \left[ c \left(a(x) +ie^{-2\phi(x)} \right)
+ d \right]$ and $c,d$ are the elements of the $SL(2,{\hbox{I}\hskip
-3.5pt\hbox{R}})$ matrix

\begin{equation}
\pmatrix{
a & b \cr
c & d \cr
} \ .
\end{equation}

\noindent Thus the norm of the gravitino and dilatino which was
calculated above is invariant under $SL(2,{\hbox{I}\hskip
-3.3pt\hbox{R}})$
symmetry and therefore
the result remains valid for the general axion-dilaton black holes of
Ref.~\cite{ADBH}.

%%%%%%%%%%%%%%%%%%%%%%%%%%%%%%%%%%%%%%%%%%%%%%%%%%%%%%%%%%%%%%%%%%%%%%

\section {Holino-hypermultiplet and rigid supersymmetry}
%{{\hskip 10pt\vskip -15pt}}

With one bound saturated, half of the original supercharges act
non-trivially.  These generate the spectrum of the resulting system.
It
is known \cite{GibbHull,Aich,Kal} to be that of the $N=2$
hypermultiplet.
In Ref.~\cite{Kal}, we called the massive $M=|Z| $ black hole
multiplet of
$N=2$ supersymmetry a holino supermultiplet.  The ERN black holes,
embedded into $N=2$ supergravity form the Clifford vacuum for the
multiplet with the highest $SU(2)$-spin $J= {1\over 2}$.  The generic
matter multiplet of $N=2$ supersymmetry is called the hypermultiplet.
There was a ``mysterious
doubling
of states" in the spectrum of the hypermultiplet, according to M.
Sohnius \cite{Son}. Indeed, the spectrum of the hypermultiplet is a
doubled
version of the massive Wess-Zumino model.
However, since the super black hole multiplets have been clearly
recognized as forming such multiplets, we now understand this
doubling.

All extreme black holes possessing  superhair have an intrinsic
way
of providing a natural doubling of the Clifford vacuum of the
corresponding multiplet.  In the basis of the supersymmetry algebra
in
which the central charge is real, there is a degeneracy of the
states:
for the same value of a mass, the charge of the black hole can take
either a positive or a negative value.  It is this degeneracy of the
black hole solutions which gives an explanation of the ``mysterious
doubling of states" in the spectrum of states with the mass, equal to
the moduli of the central charge of the state.

\begin{displaymath}
Z=  M \qquad  \leftarrow \qquad{\rm CPT- conjugate}
\qquad\rightarrow\qquad
Z=-M
\end{displaymath}

\noindent Using the superhair one can build the black hole
supermultiplet; the holino in $N=2$ supergravity.  The original
Deser-Teitelboim \cite{DesTeit} supercharge of the theory is given in
terms of the
gravitino as

\begin{equation}
Q~=~ -i{1\over{\kappa}} \gamma_5\oint_{S^2_\infty} \gamma\wedge\psi\ \ ,
\end{equation}

\noindent where ${S^2_\infty}$ is the two sphere at spatial infinity.
It was specified for the ERN black hole in \cite{GibbHull,Aich}.
Now, expanding the gravitino field in terms of the zero-mode
discussed
above and the non-zero-modes (which we ignore henceforth), we find
that
part of $Q$ is now proportional to the creation operator associated
to
the zero-mode.  This part of the supercharge generates the spectrum
of
the $N=2$ hypermultiplet.

\begin{equation}
|>_+ \;  \qquad Q_1^\dagger |>_+ \;   \qquad Q_2^\dagger |>_+ \;
\qquad Q_1^\dagger Q_2^\dagger |>_+
\end{equation}
\begin{equation}
|>_- \;  \qquad Q_1^\dagger |>_- \;   \qquad Q_2^\dagger |>_- \;
\qquad Q_1^\dagger Q_2^\dagger |>_-
\end{equation}

\noindent The upper (lower) line shows the generation of states with
the
Clifford vacuum corresponding to the positively (negatively) charged
black hole.

Before proceeding, we would like to further illustrate how the supermultiplet
of states above arises in the quantization of the
spin--$\frac{3}{2}$ field.  First we recall that the
quantization of the Rarita-Schwinger field yields all of the states
(plus parity partners) in the tensor product or Lorentz
representations:
$({1\over{2}},{1\over{2}})\otimes ({1\over{2}},0)=(1,{1\over{2}})
\oplus(0,{1\over{2}})$.  The pure spin--${1\over{2}}$ or
$({1\over{2}},0)\oplus (0,{1\over{2}})$ is projected out by imposing
$\gamma^\mu\psi_\mu=0$.  The local supersymmetry provides the gauge
parameter for this projection.  However, in the ERN background, none
of
the original supersymmetries survive as local supersymmetries.  They
act rigidly only.  This means that there are no
local
parameters which can be used to gauge away the pure
spin-${1\over{2}}$
degrees of freedom.  Thus we will be left with a dynamical Dirac
spin-${1\over{2}}$ field. Using the $N=2$ rigid supersymmetries, we then
conclude
that this spin-$1\over 2$ field is super-partnered with  bosonic fields
thereby forming the $N=2$ hypermultiplet.

The  hypermultiplet, as given by Sohnius, with {\it off-shell  central charge}
describes the
same multiplet of states with 2 complex scalars, Dirac spinor and two complex
auxiliary fields \cite{Son}

\begin{equation}
 \Phi_I = (A_I, \psi, F_I)\ \ .
\end{equation}

\noindent The underlying quantum field theory, describing the free
black
hole multiplet is
\begin{eqnarray}
{\cal L} &=& {i\over 2} (\bar \Phi^I , \delta_z \Phi_I) + {m\over 2}
(\bar \Phi^I ,
\Phi_I)\nonumber\\
 &=& {1\over{2}}
\partial_\mu A^{I \dagger}  \partial^\mu A_I~+~
i{\bar\psi}~{{\slash \hskip -6pt}{}\partial}\psi~+~ F^{I \dagger }F_I
+~ m({\textstyle\frac{i}{2}} A^{I \dagger} F_I ~-~ {\textstyle\frac{i}{2}}
F^{I \dagger } A_I ~+~ {\bar\psi}\psi)\ \ ,
\label{lagr}
\end{eqnarray}
\noindent where the central charge  transformation
\begin{equation}
\delta_z  \Phi_I = (F_I, \;  {\slash \hskip -6pt {\partial}} \psi ,
\Box A_I)\ \ ,
\end{equation}
\noindent commutes with $N=2$ global supersymmetry.

The Noether supersymmetry charge derived by quantization of the
hypermultiplet Lagrangian will generate the same set of states as the
one generated by the black hole superhair.  We may conclude therefore
that $N=2$ supergravity in the strong coupling limit may be
represented
by  soliton type states whose own dynamics may be described (before
interaction) by the free hypermultiplet action.

In the series of papers by Aichelburg
and Embacher \cite{Aich} about the supergravity solitons,  the following
conclusion has been reached.
The
free ERN black hole solitons are described by the relativistic
Lagrangian in Eq.~(6.10) of the fourth paper in Ref.~\cite{Aich}.
This
is the massive hypermultiplet Lagrangian (\ref{lagr}) with the
auxiliary fields $F_I$ excluded by their equations of motion.
Besides
explaining the hypermultiplet structure of the ERN solitons
Aichelburg
and Embacher \cite{Aich} have performed an analysis of the possible
interactions which the soliton system may have, in view of the fact
that
the multi-black hole solutions are also available.  They have made an
approximation of `` slow motion and large distance'' to find the
possible interactions in the two-soliton system.  The resulting
picture
is
the following: there are two types of solitons, with the positive and
negative charge.  The non-relativistic interaction is given in terms
of
the Hamiltonian, which is given by three parts.  One acting on
two-soliton states of both positive charges, the second one acting on
the two-soliton state of both negative charges and the last part,
acting
on the two-soliton state with solitons of opposite charges.  The
authors
suspected that the total picture may be a non-relativistic limit of
some
covariant field theory.  In such a relativistic theory, the particles
of
the opposite electric charges would become antiparticles of each
other.

Our purpose in what follows is to investigate the possible
interactions
of the ENR black holes which may be described by the full
Lorentz-covariant interacting Lagrangian whose free part is the
hypermultiplet action with the off-shell central charge (\ref{lagr}).
It is important to stress that the relativistic action (\ref{lagr})
describes both types of soliton states with positive and negative
charges
and, in this respect, is capable of representing these two types of
non-relativistic solitons as antiparticles of each other.

Having seen that the $N=2$ hypermultiplet arises in the quantization of
the zero-mode part of the gravitino, we now wonder if such a
multiplet
may be placed on the background for which this zero-mode exists.  To
check this, we must first find rigid parameters. {}From the structure
of
the multiplet, we see that we need two such parameters.

Fix the masses of a ERN multi--black hole background.
Identify the associated
Killing and anti-Killing spinors. For given masses, these spinors are
distinguished by different signs of the charges;
call these two parameters collectively,
$\epsilon^I$.  Now place the $N=2$ massive
hypermultiplet on this
background.  It is important that all derivatives (covariant with
respect to this background) may be replaced by ${{\slash \hskip
-8pt}{}
\nabla}$'s at the expense of a surface term.  Consequently, since the
$\epsilon$'s are constant with respect to ${{\slash \hskip -8pt}{}
\nabla}={{\slash \hskip -8pt}{} {\hat\nabla}}$, the action

\begin{eqnarray}
S^{N=2}_{\rm hyp}~=~ \int d^4x&\sqrt{-g}&\Bigl[ {1\over{2}}
g^{\mu\nu}\nabla_\mu A^{I \dagger} \nabla_\nu A_I~+~
i{\bar\psi}~{{\slash \hskip -8pt}{}\nabla}\psi~+~ F^{I \dagger} F_I
\nonumber \\
\relax
&&+ m({\textstyle\frac{i}{2}} A^{I \dagger}  F_I ~-~ {\textstyle\frac{i}{2}}
F^{I \dagger} A_I ~+~ {\bar\psi}\psi)\Bigr]\
\label{N2ACT}
\end{eqnarray}

\noindent  is invariant under the $N=2$ rigid
supersymmetry transformations,
\begin{eqnarray}
\delta A_I&~=~& 2{\bar\epsilon}_I \psi~+~ \zeta F_I\ \ ,
\nonumber \\ \relax
\nonumber \\ \relax
\delta\psi &~=~& -i{\epsilon}^IF_I ~-~
i~{{\slash \hskip -8pt}{}\nabla}\epsilon^IA_I
{}~+~ \zeta ~{{\slash \hskip -8pt}{} \nabla}\psi\ \ ,
\nonumber \\ \relax
\nonumber \\ \relax
\delta F_I &~=~& 2{\bar\epsilon}_I
{{\slash \hskip -8pt}{} \nabla}\psi~+~ \zeta
\triangle A_I \ \ .
\end{eqnarray}
The parameters, $\epsilon_I$ may be combined to form the Dirac spinor
\begin{equation}
\epsilon (x) ~=~ V^{-1/2}(x) \; \epsilon_0 ~=~\left  (1\ +\ \sum_{s}
\frac{G
M_s}{|\vec
x\ -\ \vec
x_s|}\right )^{-1/2} \epsilon_0\ \ ,
\end{equation}

\noindent and central charge parameter is

\begin{equation}
\zeta (x) = V^{-1} \zeta_0 = \left (1\ +\ \sum_{s} \frac{G M_s}{|\vec x\ -\
\vec
x_s|}\right )^{-1} \zeta_0 \ ,
\end{equation}

\noindent where $\epsilon_0, \zeta_0$ are the values of the global
supersymmetry and central charge transformation parameters in the
flat
background.

The replacement of ${{\slash \hskip -8pt}{} \nabla}$ by
${\slash \hskip -8pt}{} {\hat\nabla}$ was made so that the
(anti-)Killing
spinors  may be used thereby allowing
us
to establish these supersymmetries.
The rigid $N=2$ supersymmetry elucidated above is  based on the
parameter of
transformation which solves a massless  Dirac equation in the
background of the
Majumdar--Papapetrou  metric.  One can show that
 \begin{equation}
\epsilon(\vec x)~=~ V^{-1/2}\epsilon _0\ \ ,
\end{equation}
are solutions of the Dirac equation
\begin{equation}
{{\slash \hskip -8pt}{} \nabla}\epsilon~=~ 0\ \ .
\end{equation}
Here, $\epsilon _0$ is an arbitrary constant Dirac spinor.

The supersymmetry algebra reads
\begin{eqnarray}
\{{\bar Q}^I, Q_J\} &~=~& i2 \delta_J{}^I
({{\slash \hskip -8pt}{} {\hat\nabla}} {}~+~ {\cal Z})\
\ ,\nonumber \\ \relax
\nonumber \\ \relax
[{\cal Z},Q_I]&~=~&0\ \ .
\end{eqnarray}
 where $Q_I$ is the supersymmetry charge and $\cal Z$ is the
central charge generator.  We note that the parameters which appear
on
the right-hand-side of the commutator of two supersymmetries are
$K_\mu={\bar\epsilon}_I^{(2)}\gamma_\mu\epsilon^{(1)\,I}$ and
$\zeta={\bar\epsilon}_I^{(2)}\epsilon^{(1)\,I}$.  The first is a
Killing vector
while the second is identified as the central charge parameter.

Thus, the presence of gravitino zero modes
and rigid supersymmetry in a certain curved background has led us,
following a brief analysis of its quantization, to an action which is
that of a novel rigidly supersymmetric theory.  This action is
presumably
the candidate action for
the supersymmetric excitations of the ERN black hole.

%%%%%%%%%%%%%%%%%%%%%%%%%%%%%%%%%%%%%%%%%%%%%%%%%%%%%%%%%%%%%%%%%%%%%%

\section{Discussion}
%{{\hskip 10pt\vskip -15pt}}

We have found that the normalizability of the gravitini zero modes
is correlated with the existence  of a  modified WIN construction in the
absence of the
source term.
The evaluation of the integrals for the norms of  the
black
holes, which we have performed in this paper was consistent with the
 use of the Gauss theorem with the contribution coming only from the
surface at
asymptotic infinity.  The reason for this was the vanishing of the
integrand at
the horizon.
If the constant-time slices include singularities, this
term may contribute.   It would be interesting to
know what happens in the more complicated case of the supersymmetric
but singular IWP
configurations of $N=2$ supergravity \cite{Tod} and in the case of
the
supersymmetric IWP configurations of $N=4$ supergravity
(dilaton-axion
gravity) \cite{KKOT}.

Our results for the finiteness of the gravitino norm in the $3+1$ dimensional
MP
configurations and axion-dilaton black holes are in contrast with the
situation in
$2+1$ dimensions studied in \cite{BecStr}, where the norm was found to be
infinite. Additionally, it is interesting that in a closely related
$2+1$ dimensional theory it has been recently found \cite{Howe}
 that no bound, which is normally derived from the standard WIN construction,
exists.

We would  like  now to compare our calculations of the norm with the
calculations
of the
moduli space of the two-black-hole configurations \cite{FE}. Some of
the
integrals used there resemble the integrals we have found for the
norms of the
fermion zero modes. In
particular, for the black holes
considered here, the expression used for the moduli space metric was
given in Ref.~\cite{FE}. For the ERN case
 ($a=0$)  and for dilaton black holes ($a=1$) the moduli space metric
was
calculated from the following integrals:
\begin{eqnarray}
\gamma(r;a) &\sim &\int d^3x~ V^{2(1-a^2)\over (1+a^2)} \left(|\vec \partial
V|^2 -
{m_1^2
\over
|\vec r_1|^4}-
{m_2^2 \over |\vec r_2|^4}\right) \ , \label{mod}\end{eqnarray}
where
\begin{equation}
V = 1 +   {m_1 \over |\vec r_1|} +
{m_2 \over |\vec r_2|} \ ,   \qquad  \vec \partial V =  {m_1 \vec r_1 \over
|\vec r_1|^3} +
{m_2 \vec r_2 \over |\vec r_2|^3} \ ,
\end{equation}
and $\vec r_1 = \vec x - \vec x_1$, $\vec r_2 = \vec x - \vec x_2$, $\vec r =
\vec x_1 - \vec x_2$.
The crucial difference between these expressions and
our expression for the gravitini norm in the two-black
hole case
\begin{equation}
\int d^3x~ V^{-2} |\vec \partial V|^2\ \ ,
\end{equation}
%from  both  expressions in (\ref{mod})
is  the pre-factor  $V^{-2}$. For the moduli
metric
such terms are $V^{+2}$ or $V^{0}$.
If we take the domain of integration to be ${\hbox{I}\hskip
-3.3pt\hbox{R}\hskip -2pt}^3$, as was done in
Ref.~\cite{FE},
near each horizon $V^{-2}
\rightarrow 0 $,
however $V^{+2} \rightarrow \infty $ and $V^{0} \rightarrow 1$. We
understand
therefore that the calculations of the moduli space metric may not be
unambiguous and may require  additional confirmation. The choice
of
regularization near the horizon may, under some circumstances, affect
the result.

It is therefore quite satisfying that the expressions for the
fermion zero-mode norms for
all black holes which we have considered in this paper were
particularly simple.  In particular, if we were to extend the domain
of
integration to ${\hbox{I}\hskip -3.3pt\hbox{R}\hskip -2pt}^3$, we
would
not need to introduce any regularization near
the black
hole horizons. However, our considerations apply only to
supersymmetric black
holes, whereas
the moduli space metric has  divergences near the horizon
for
arbitrary dilaton coupling $a$. The zero mode calculation which we
have
performed would not be generalized for arbitrary dilaton coupling.
The
importance of the finiteness
of the norm lies in the fact that this allows us to construct the
supersymmetric multiplets including the black hole partners. This
presents an
alternative possibility to study black hole  supersymmetric
multiplets and
their possible interactions in  the framework of relativistic quantum
field
theory or string theory or perhaps even string field theory, avoiding
the
non-relativistic approximation.

In the present paper, we have argued that the dynamics of the
non-interacting
supersymmetric holino multiplet with the bosonic part given by the
ERN black
hole
is described by the free Sohnius hypermultiplet action with an
off-shell
central charge.
The BPS $M=|Z|$ condition is  realized only on-shell. We have shown
that this
theory can be placed in the corresponding
gravitational multi-black-hole background with the global
supersymmetry of the
free theory generalized to the rigid one in the background. Under the
condition
that the most general interaction of these super-black-hole states
preserves the $N=2$ supersymmetry with the central charge equal to
the mass of
the multiplet on shell, one can try to describe the interacting  ERN
black
holes in the framework of a relativistic quantum field theory.  We
expect that
such a description would make use of  the recent
progress in understanding the structure of the superpotential for
$N=2$ supersymmetric sigma models \cite{GA}.

%%%%%%%%%%%%%%%%%%%%%%%%%%%%%%%%%%%%%%%%%%%%%%%%%%%%%%%%%%%%%%%%%%%%%%

\section*{Acknowledgements}
Our gratitude is expressed to E. Bergshoeff for  fruitful discussions regarding
this work.
We would also like to acknowledge the useful discussions on the moduli
space metric we have had with G. Gibbons, D. Eardley and K.
Shiraishi. We have learned about the recent progress with   $N=2$
supersymmetric superpotential  from  V. Ogievetsky and A. Galperin.
R.B.~and T.O.~would like to express their gratitude to the Physics
Department of Stanford University for its hospitality and financial
support.  The work of R.B. was supported in part by funds provided by
the U.S. Department of Energy (D.O.E.) under cooperative agreement
\# DE-FC02-94ER40818.
The work of R.K.~was supported by the NSF grant
PHY-8612280.
The work of T.O.~was also supported by a European Union {\it Human
Capital and Mobility} program grant.

%%%%%%%%%%%%%%%%%%%%%%%%%%%%%%%%%%%%%%%%%%%%%%%%%%%%%%%%%%%%%%%%%%%%%%

%

\end{document}